\documentclass[prb,aps,amsmath,amssymb,amsfonts,floatfix,groupedaddress,showpacs,twocolumn]{revtex4-1}
\usepackage{longtable}
\usepackage{graphicx}
\usepackage{epsfig}
\usepackage[dvips]{color}
\usepackage{dcolumn}
\usepackage{nicefrac}

\newcommand{\bk}{\mathbf{k}}

\def\beq{\begin{equation}}
\def\eeq{\end{equation}}
\def\be{\begin{equation}}
\def\ee{\end{equation}}

\def\t{\mbox{tr}\,}
\def\cG0{{\cal G}_0}
\def\cG{{\cal G}}

\def\a{\alpha}

\def\uc2{$U_{c2}$}
\def\uc1{$U_{c1}$}
\def\bavs3{BaVS$_3$}

\def\t2g{$t_{2g}$}

\def\a1g{$a_{1g}$}

\begin{document}

\title{Impact of the Dzyaloshinskii-Moriya interaction
in strongly correlated itinerant systems}
\author{Sergej Schuwalow}
\affiliation{I. Institut f{\"u}r Theoretische Physik,
Universit{\"a}t Hamburg, D-20355 Hamburg, Germany}
\author{Christoph Piefke}
\affiliation{I. Institut f{\"u}r Theoretische Physik,
Universit{\"a}t Hamburg, D-20355 Hamburg, Germany}
\author{Frank Lechermann}
\affiliation{I. Institut f{\"u}r Theoretische Physik, 
Universit{\"a}t Hamburg, D-20355 Hamburg, Germany}

\begin{abstract}
Spin-only approaches to anisotropic effects in strongly interacting materials are
often insufficient for systems close to the Mott regime. Within a model context, 
here the consequences of the low-symmetry relevant Dzyaloshinskii-Moriya (DM) 
interaction are studied for strongly correlated, but overall itinerant, systems. 
Namely, we investigate the Hubbard bilayer model supplemented by a DM term at half 
filling and in the hole-doped regime. As an add-on, further results for the two-
impurity Anderson model with DM interaction are also provided. The model 
Hamiltonians are treated by means of the rotational invariant slave boson technique 
at saddle point within a (cellular) cluster approach. Already small values of the 
anisotropic interaction prove to have a strong influence on the phases and 
correlation functions with increasing $U$. An intriguing metallic spin-flop phase 
is found in the doped bilayer model and a reduction of the RKKY exchange in the
two-impurity model.    
\end{abstract}

\maketitle

\section{Introduction}
The effect of anisotropic magnetic exchange on the atomistic
level has been recently brought back to a centre of interest in condensed matter 
physics due to its intriguing importance in e.g. the search for multiferroic 
materials~\cite{ede05,kho09}, the understanding of complex metallic magnetic surface 
structures~\cite{bod07} or the phenomenology of topological 
insulators.~\cite{zhu11} A hallmark step in this research topic
has already been performed some fifty years ago by Dzyaloshinskii~\cite{dzy58} and 
Moriya,~\cite{mor60} who derived an effective spin-spin interaction term from the 
spin-orbit coupling in low-symmetry cases, the so-called Dzyaloshinskii-Moriya 
(DM) interaction. 
The DM term or more generic magnetic anisotropies are nowadays believed to play 
furthermore a prominent role in many strongly correlated materials. However nearly 
exclusively, theoretical studies in this context were in the past bound to pure spin 
models without itinerancy, leaving the impact of charge fluctuations aside. Yet the
latter are surely important, e.g. close to the Mott-critical regime of the 
metal-insulator transition. Allying the Hubbard model with spin-orbit terms has 
just recently gained rising interest.~\cite{pes10,men10}

In the present work we aim at a minimal modeling of the influence of the DM 
interaction in the strongly correlated metallic regime. There are many  
specific materials problems motivating such a case study, namely the complex 
magnetic behavior of doped cuprate systems,~\cite{thi88,cof91,jur06} 
manganites,~\cite{hir98,mit01} and mono-oxides~\cite{sol98} as well as anisotropic 
magnetic effects close to the metal-insulator transition in low-dimensional organic 
compounds~\cite{kag08} or in the context of transition-metal impurities on metallic 
surfaces.~\cite{zho10}
While standard direct and indirect exchange processes favor collinear alignment of 
the local spins generated in the strongly correlated metallic regime, the DM 
interaction tends to align the spins in a perpendicular fashion. Thus the competition
between the former conventional exchange processes and the DM interaction 
within an itinerant system shall give rise to nontrivial physics resulting in 
sophisticated spin arrangements/orderings. 

To keep things simple and to build up on a somewhat canonical approach, we rely on 
two basic models, namely the bilayer model of two coupled single-band
Hubbard planes~\cite{zie96,moe99,fuh06,kan07,lec07,bou08,haf09,yos09} and the 
two-impurity Anderson model.~\cite{jay82,fye87,jonkot89,fye89,schi96,nis06,ferr09} 
The former Hamiltonian allows for a DM coupling between two lattice planes in
the thermodynamic limit, whereas the latter one provides the possiblity to study 
the DM term within a local perspective via interacting impurities coupled to the 
same bath. Both setups render it possible to investigate
nearest-neighbor (NN) correlation functions between sites in an 
itinerant background. Of course, such modelings are not sufficient to grasp the 
very details of the above named materials problems, yet it will be shown that the 
computed phenomenology is far from trivial and may apply to generic realistic 
phenomena. One key focus in the context of the Hubbard bilayer lattice is thereby on 
the competition between the antiferromagnetic (AFM) tendencies driven by direct 
exchange and the DM term within the metallic state. It will become clear that 
already rather small values of the DM integral may have a significant influence on 
the magnetic ordering tendencies in the larger Hubbard $U$ range, i.e., the AFM 
state is rather sensitive to only minor DM perturbations. A rich phase diagram 
results from the interplay of kinetic energy, onsite Coulomb and DM interaction.
The latter also has important consequences in the two-impurity model, where its
favor for perpendicular spin arrangement severely affects the local-limit 
competition between singlet-forming Kondo-screening and triplet-forming 
Ruderman-Kittel-Kasuya-Yosida (RKKY) interaction.

In the following we define the model Hamiltonians as well as our mean-field 
approach in section~\ref{sec:mod}. The results for the Hubbard bilayer 
at half filling and in the hole-doped case are discussed in section~\ref{sec:hubi}.
Some basic observations retrieved from the studies on the two-impurity Anderson 
model with DM interaction will be presented in section~\ref{sec:tiam}.

\section{Hamiltonians and Theoretical Approach\label{sec:mod}}
\begin{figure}[b]
\centering
\includegraphics*[width=7cm]{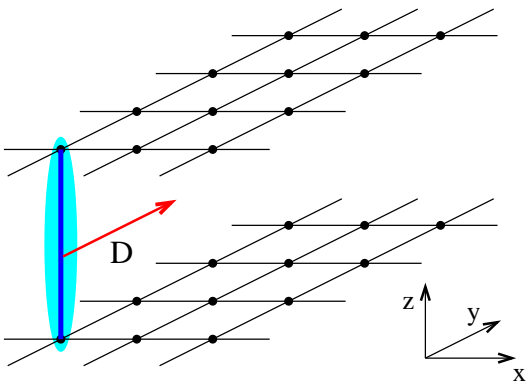}%
\caption{(Color online) Bilayer model with DM interaction. The vertical blue line
is a representant of the inter-layer hopping $t_\perp$, which of course is applied
at every lattice point, and the ellipse marks the two-site cluster. The
DM integral vector ${\bf D}$ is chosen to point in $y$ direction.
\label{fig:mod}}
\end{figure}

The first problem addressed here consists of two coupled two-dimensional 
infinite square-lattice planes with one orbital per site each facing an on-site Coulomb 
repulsion $U$ (see Fig.~\ref{fig:mod}). In both planes the electron 
dispersion is defined by identical simple NN hopping $t$. The inter-plane coupling 
is realized via a perpendicular hopping $t_\perp$ as well as a DM interaction 
mediated by the vector integral ${\bf D}$. The model Hamiltonian is accordingly 
written as
\begin{eqnarray}\label{eq:biham}
H_{\rm BL}=&&-t\sum_{\stackrel{\alpha\sigma}{\langle i,j\rangle}}
(c^\dagger_{\alpha i\sigma}c^{\hfill}_{\alpha j\sigma}+{\rm h.c.})+
t_{\perp}\sum_{i\sigma}(c^\dagger_{1i\sigma}c^{\hfill}_{2i\sigma}
+{\rm h.c.})\nonumber\\
&&+\;U\sum_{\alpha i}n_{\alpha i\uparrow}n_{\alpha i\downarrow}+
\sum_i{\bf D}\cdot({\bf S}_{1i}\times{\bf S}_{2i})\quad,
\end{eqnarray}
where $c^{(\dagger)}_{\alpha i\sigma}$ creates/annihilates an electron in layer
$\alpha$=1,2 at lattice site $i$ with spin projection $\sigma$=$\uparrow,\downarrow$.
The $\nu$=$x,y,z$ component of the spin operator at each site $i$ of an 
individual layer $\alpha$ is provided by
$S^{(\nu)}_{\alpha i}$=$1/2\,c^\dagger_{\alpha i\sigma}\,\tau^{(\nu)}_{\sigma\sigma'}\,c^{\hfill}_{\alpha i\sigma'}$ with the Pauli matrices $\tau^{(\nu)}$. In general, 
the vector interaction ${\bf D}$ is defined perpendicular to the bond between the 
involved lattice sites.~\cite{dzy58,mor60} Since otherwise there is a freedom of 
choice for the explicit direction, we pick ${\bf D}$ to point along the $y$ axis,
i.e. ${\bf D}$=$D\,{\bf e}_y$. 
Note that the DM interaction may only occur if the inversion symmetry is broken. To 
facilitate this in the present case, one could e.g. think of an inter-layer coupling 
originally established via oxygen with an angle deviating from 180$^\circ$. 

In the smaller second part of this paper, we take the opportunity to also briefly 
discuss the well-known two-impurity Anderson model (TIAM) supplemented by a DM
interaction between the impurities. We write that model in the form
\begin{eqnarray}\label{eq:impham}
H_{\rm TIAM}=&&\sum_{\bk\sigma}\varepsilon_{\bk}
c^\dagger_{\bk\sigma}c^{\hfill}_{\bk\sigma}+
\varepsilon_d\sum_{i\sigma}n_{i\sigma}\nonumber\\
&&+\,V\sum_{\bk i\sigma}(c^\dagger_{\bk\sigma}d^{\hfill}_{i\sigma}+{\rm h.c.})
+\;t_{12}\sum_{\sigma}(d^\dagger_{1\sigma}d^{\hfill}_{2\sigma}
+{\rm h.c.})\nonumber\\
&&+\,U\sum_{i}n_{i\uparrow}n_{i\downarrow}+
{\bf D}\cdot({\bf S}_{1}\times{\bf S}_{2})\quad,
\end{eqnarray}
with the impurity-electron operators $d^{(\dagger)}_{i\sigma}$ ($i$=1,2), the 
impurity-level energy $\varepsilon_d$ and the impurity-impurity hopping $t_{12}$. 
The bath has associated operators $c^{(\dagger)}_{i\sigma}$ and a dispersion 
$\varepsilon_{\bk}$. The impurity-bath coupling is denoted by $V$ and the Hubbard 
$U$ is located on the impurities with $n_{i\sigma}$=$d^{\dagger}_{i\sigma}d^{\hfill}_{i\sigma}$. In the present approach the bath is treated explicitly through a 
three-dimensional simple-cubic dispersion with bandwidth $W$=$12t$, choosing $t$=0.5.
For the direction of ${\bf D}$ again the $y$ axis is selected. The impurities have a 
common bath, yet $V$ is assumed here to be $k$-independent and the explicit 
impurity-impurity distance is formally set to zero. A constant value of $V$=$-0.5$
is chosen in the present work. Hence only the local part of the RKKY interaction is 
accessible. Such a modeling is e.g. important for understanding the local spin 
interactions between correlated atoms on metallic surfaces,~\cite{zho10} where 
there is indeed an intriguing interplay between conventional direct exchange, RKKY 
interaction, Kondo effect and anisotropic exchange.

For the numerical solution of the model Hamiltonians discussed here, the 
rotationally invariant slave-boson (RISB) formalism~\cite{li89,lec07} in the 
saddle-point approximation, similar to the generalized Gutzwiller 
approach,~\cite{bue98} is employed. The RISB methodology amounts to a decomposition 
of an electron operator $a^{\hfill}_{\nu\sigma}$ with generic orbital/site index 
$\mu$ via $\underline{a}_{\mu\sigma}$=$\hat{R}[\phi]^{\sigma\sigma'}_{\mu\mu'}f_{\mu'\sigma'}$ into its quasiparticle (QP) part $f_{\mu\sigma}$ and the remaining 
high-energy excitations carried by the set of slave bosons $\{\phi_{An}\}$. Here 
$A$ denotes a chosen localized basis state and $n$ relates to the given QP degree 
of freedom. Two constraints, the first enforcing the normalization of the bosonic 
content and the second keeping an eye on the match of the bosonic and the fermionic 
occupation matrix, are established on site-average at saddle-point through the 
Lagrange-multiplier matrix $\Lambda$.~\cite{lec07} In order to describe 
inter-atomic correlations adequately, a two-site (cellular-cluster) framework is
used. This cluster connects two NN lattice sites between the layers in the Hubbard
bilayer and the two impurities in the TIAM. It amounts to a local cluster approach 
to the electronic self-energy, whereby $\Sigma_{12}(\omega)$ 
incorporates terms linear in frequency as well as static 
renormalizations.~\cite{lec07} Therewith the low-energy behavior may be adequately 
expressed and inter-site correlation functions as well as multiplet
weights on the cluster can be retrieved. Importantly, the formalism allows for 
full spin and orbital rotational invariance, needed to account for the competition 
between isotropic and anisotropic interactions. In this respect the slave bosons 
may become true complex numbers and $\Lambda$ can be expanded via Pauli matrices in 
each orbital sector (with allowed off-diagonal terms between these sectors). Albeit 
the calculations are formally performed at temperature $T$=0, a small gaussian
smearing for the k-point integration introduces a minor $T$ scale.  For this reason 
the energetics are discussed in terms of the free energy $F$. Note that in the
numerical solution of the TIAM, a three-orbital model is effectively treated within
RISB, whereby the bath enters through its band dispersion. Thus the bath degrees of 
freedom are not integrated out, but are handled explicitly. In principle, a 
correlated-bath scenario may also be studied, however we here always keep 
$U_{\rm bath}$=0. Nevertheless, correlation effects are introduced within the
bath due to the coupling to the correlated impurities. The investigated half-filled
scenario of the model is either achieved by setting $\varepsilon_d$=$-U/2$ or 
through an additional Lagrange multiplier fixing the electron occupation on the 
bath according to the total filling $N$=3.

\section{ Hubbard bilayer model\label{sec:hubi}}
The original Hubbard bilayer without DM interaction has already been addressed in
several works,~\cite{zie96,moe99,fuh06,kan07,lec07,bou08,haf09,yos09} most often 
concerning the electronic phase diagram when varying the ratio $t_\perp/t$. Here 
however the main interest lies on the ratio $U/|{\bf D}|$ for the coupled 
square-lattice layers with bandwidth $W$=$8t$. In the following, we restrict the
discussion to cases $t_\perp/t$$<$1 with all the energies given in units of the half 
bandwidth $4t$.

Concerning the electronic phases studied within the current mean-field approach, we 
restrict the discussion to local cluster orderings, i.e., neglect long-range order 
parameters suitable for e.g. spin spirals. Such more intricate instabilities are 
planned to be addressed in more concrete materials-connected future modelings. Here 
the focus is first on the interplay of the fundamental short-range correlation 
processes in the strongly correlated metallic regime. Note however that in the 
present context the cluster description does not account for intra-layer inter-site 
self-energies.

\subsection{Half-filled case}
At half filling, each layer accomodates one electron and the whole system is
therefore susceptible to a Mott transition. We study two cases, namely the one
of weakly coupled layers $(t_\perp$=$0.025)$ and the other with stronger
inter-layer hopping $(t_\perp$=$0.1)$.

\begin{figure}[t]
\centering
\includegraphics*[width=8.5cm]{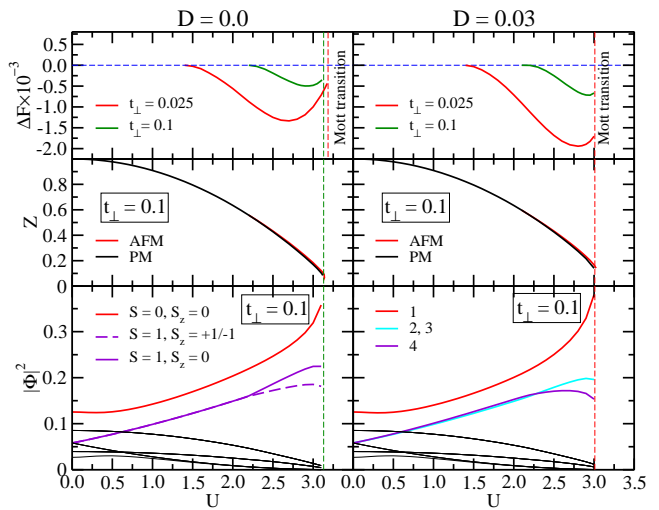}%
\caption{(Color online) Energetics, QP weight and multiplet weights with
increasing $U$ for the bilayer model at half filling for two values of $t_\perp$ and 
$D$, respectively. Free energies are normalized to the one of the PM phase.
The thick lines in the multiplet-weights plot correspond to the
states in the two-particle sector.
\label{fig:hf1}}
\end{figure}

Figure~\ref{fig:hf1} shows the phase competition within the half-filled model with 
increasing the Hubbard $U$. The computations allow for the stabilization of two 
metallic phases, namely the paramagnetic (PM) and antiferromagnetic-between-layers 
(AFM) ones.         
\begin{figure}[t]
\centering
\includegraphics*[width=8.5cm]{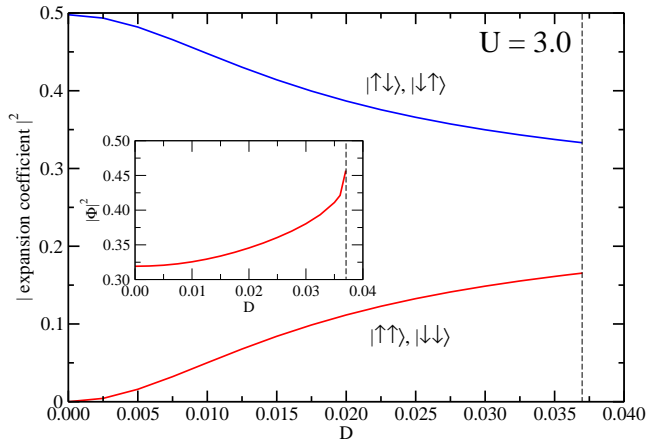}%
\caption{(Color online) Evolution of the Fock-state contributions to the local state 
with the highest slave-boson amplitude (compare with Fig.~\ref{fig:hf1}) for 
$U$=3. Note that for $D$$\neq$0 this state "1" is no longer an eigenstate of 
the $S^2$ and the $S_z$ operator. The inset exhibits the $D$ dependence of the 
state.\label{fig:hfmulti}}
\end{figure}
\begin{figure}[b]
\centering
\includegraphics*[width=8.5cm]{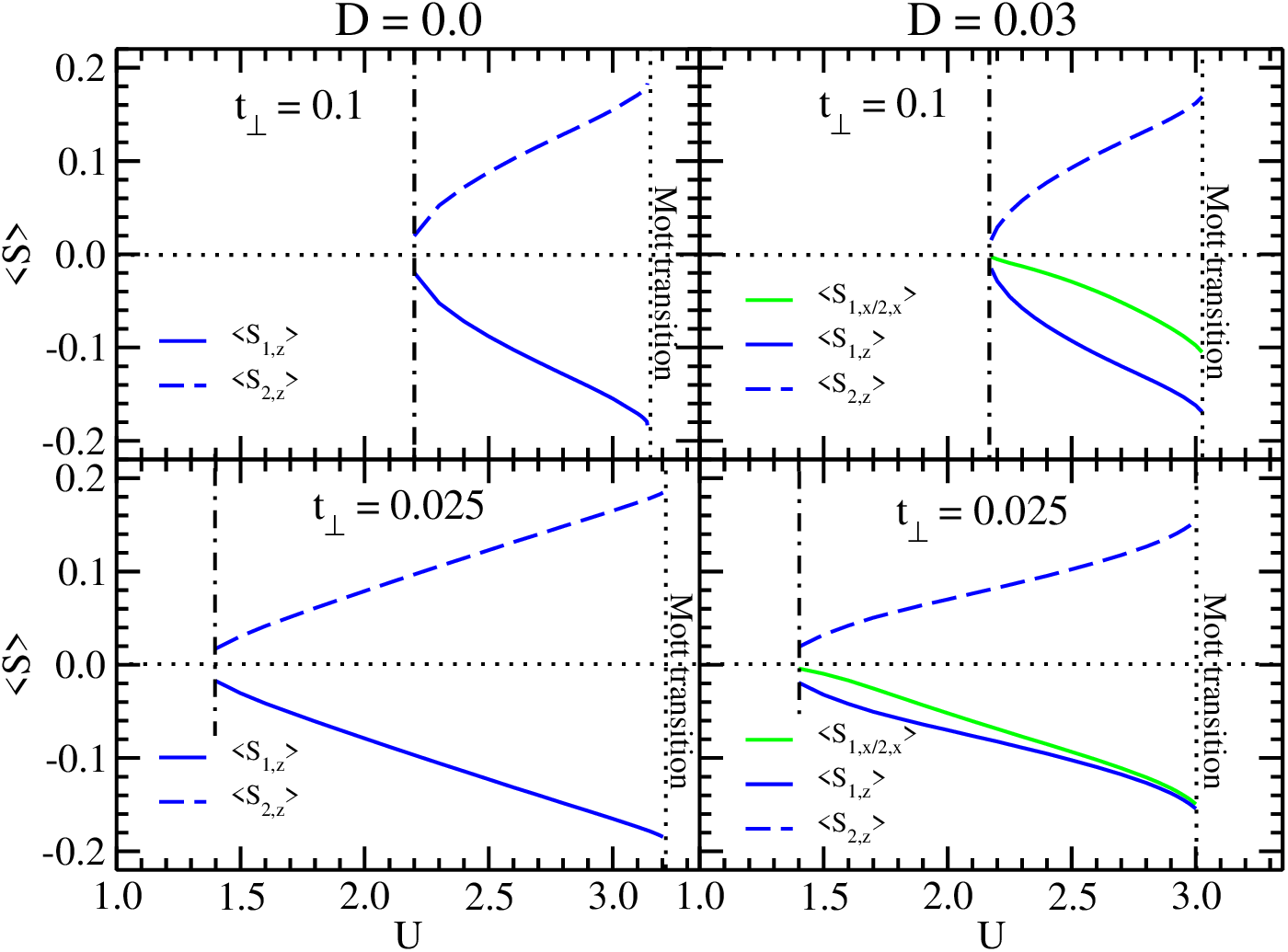}%
\caption{(Color online) Interaction dependence of the spin expectation values in the 
AFM phase. Nonzero values of the DM coupling introduce a spin component pointing 
along the $x$ axis (due to the choice for the direction of the $D$ vector). The
vertical dot-dashed lines mark the AFM transition point.
\label{fig:hf2}}
\includegraphics*[width=8.5cm]{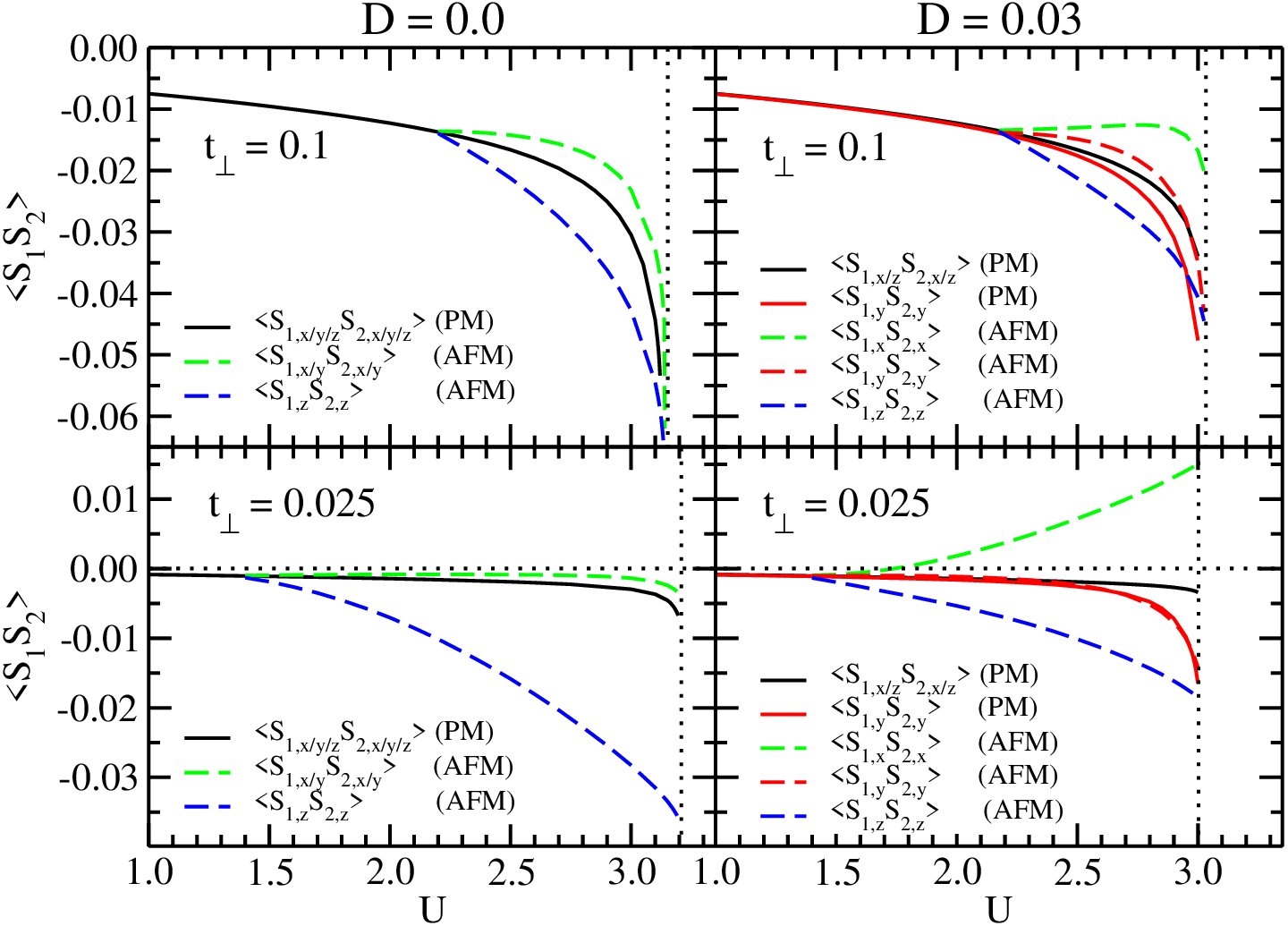}%
\caption{(Color online) Same as Fig.~\ref{fig:hf2}, here for the spin-spin 
correlation functions between the layers. Solid lines: PM phase, dashed lines: 
AFM phase.\label{fig:hf3}}
\end{figure}
From the inspection of the free-energy differences it is nonsurprisingly seen that
in general the AFM phase wins over the PM phase at larger $U$. Thereby a smaller 
$t_\perp$, and hence a smaller bonding/anti-bonding splitting, supports the 
building-up of the AFM phase, in line with DMFT calculations employing quantum 
Monte-Carlo solvers for the impurity problem.~\cite{kan07,haf09} A further 
gain in AFM free energy is observed at fixed $U$ when introducing the DM 
interaction, but with only marginal shifts of the phase onset towards smaller $U$. 
The difference between the two critical $U$=$U_c$ for the two different $t_\perp$ vanishes
 with $D$, while in the case of $D$=0 the $U_c$ for $t_\perp$=0.1 is clearly smaller. However 
the general evolution of the QP weight 
$Z$=$(1-\partial\Sigma/\partial\omega)^{-1}|_{\omega=0}$ with $U$ does not display 
strong changes with the introduced anisotropic interaction.

In addition, Fig.~\ref{fig:hfmulti} also displays the slave-boson weight of the 
local multiplet state that dominates at half filling in the relevant two-particle 
sector. For $D$=0 the bilayer Hamiltonian (\ref{eq:biham}) 
commutes with $\{S^2,S_z\}$ and thus singlet and triplet states form the local 
two-particle eigenbasis. As expected, the singlet remains strongest up to the Mott 
transition, followed by the triplet states (whith their degeneracy lifted when 
entering the AFM phase). In the case of finite $D$, the picture formally looks very 
similar, but Hamiltonian and spin operators are no longer commuting operators
and the respective two-particle states thus are not true spin eigenfunctions. 
That is easily understood from the DM interaction favoring a perpendicular alignment 
of the local spins, contrary to the originally preferred collinear states. It is 
nicely illustrated in Fig.~\ref{fig:hfmulti}, where the inter-mixing of the singlet 
and triplet Fock-state building blocks with growing $D$ is exhibited. 

Figure~\ref{fig:hf2} shows the evolution of the spin moments in the two layers
with increasing $U$. For $D$=0 only $\langle S_z\rangle$ adopts a nonzero value in 
the AFM phase, with a steeper increase for larger $t_\perp$. However with finite 
$D$ also a sizeable $x$ component of $\langle S\rangle$ shows up and grows until 
 $U_c$ is reached. For the smaller $t_\perp$=0.025 the value for 
$\langle S_x\rangle$ even equals the corresponding $\langle S_z\rangle$ magnitude. 
A lower $t_\perp$ apparently also effectively increases the relative 
tendency towards the non-collinear spin alignment driven by the DM coupling. Note 
that the DM interaction not only modifies the AFM phase, but has an impact in 
the PM state as well. There $\langle S_{1,y}S_{2,y}\rangle$ exhibits more AFM-like 
character and the corresponding $(x,z)$ correlation functions show minor weakened 
AFM-like tendencies, both compared to the $D$=0 case.
Close to the Mott transition the larger $t_\perp$ results in a stronger (coherent) 
spin response for $D$=0, as retrieved from the inter-layer spin correlation 
functions plotted in Fig.~\ref{fig:hf3}. For nonzero $D$ the correlation between 
the $x$ components, i.e. $\langle S_{1,x}S_{2,x}\rangle$, appears to behave 
especially more disconnected from the $z$ component for the smaller $t_\perp$.  

\begin{figure}[t]
\centering
\includegraphics*[width=8.5cm]{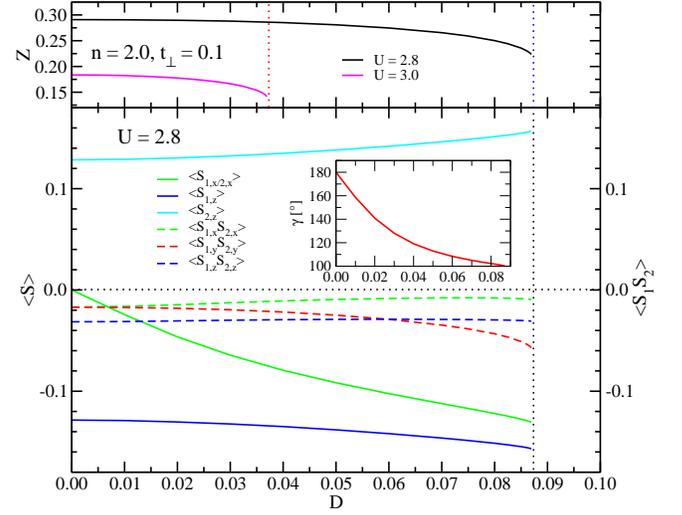}%
\caption{(Color online) Influence of the DM coupling on the QP weights, 
spin moments and spin-spin correlation functions for the AFM half-filled bilayer 
model. The inset shows the evolution of the angle between the spin moments in the 
two layers.\label{fig:hf4}}
\end{figure}
In order to gain further insight into the impact of the DM term, Fig.~\ref{fig:hf4} 
depicts explicitly the $D$ dependence for fixed $U$. The Mott transition itself 
may be tuned over a rather wide range of the anisotropic interaction. Whereas the 
spin moment in the $x$ 
direction shows a strong variation with $D$, the spin-spin correlations are only 
weakly dependent thereon. Albeit no resulting $\langle S_y\rangle$ value exists, the 
correlations along $y$ still appear to gain strongest in magnitude from a larger $D$.
It is also visualized that the angle $\gamma$ between the local spins on the
adjacent layers indeed increasingly deviates from the AFM-ideal 180$^{\circ}$ with
growing DM interaction. Close to the Mott transition, the value 
$\gamma$$\sim$100$^{\circ}$ is nearby the DM-ideal value of 90$^{\circ}$.

\subsection{Hole-doped case}

We now turn to the effects of doping the bilayer model away from half filling. 
For investigating the simultaneous effects of doping, on-site Coulomb interaction 
and inter-site DM interaction we set $t_\perp$=0.1 and first fix the Hubbard 
interaction to $U$=3. As can be seen from Fig.\ref{fig:hf1} the value of $U$
puts the system just below the Mott transition at half filling, i.e. strong 
correlations with the quasiparticle weight $Z$$\sim$0.2 exist. 

\begin{figure}[b]
\centering
\includegraphics*[width=8.5cm]{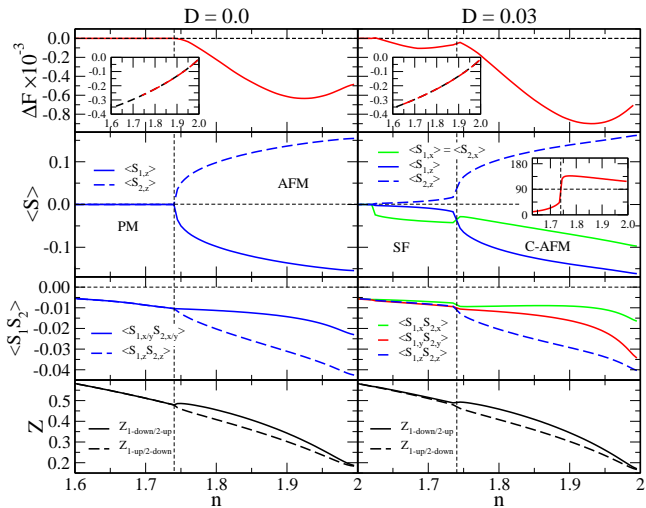}%
\caption{(Color online) Influence of hole doping on the bilayer model, 
with and without DM interaction for $t_\perp$=0.1 and $U$=3. Insets at the top 
show the free-energy curve, with the region where AFM order is (meta)stable marked 
in red. C-AFM marks the canted antiferromagnetic and SF the spin-flop phase. The 
inset in the right-middle depicts the evolution of the angle $\gamma$ between spins.
\label{fig:hd1}}
\end{figure}

The results of hole doping $\delta$=$2$$-$$n$ for the system in the filling range 
$n\in [1.6,2.0]$ are summarized in Fig.~\ref{fig:hd1}. Let us first discuss the $D$=0 
case. Starting from half filling, the system is in the AFM phase for the chosen $U$ 
value. With increasing $\delta$ the order parameter $\langle S_z\rangle$ decreases,
until it vanishes close to $n$=1.74 and the PM phase sets in (at reduced 
spin-spin correlations and larger QP weight). When including a DM interaction with
$D$=0.03 in the model, the situation becomes more intriguing. Again the AFM (C-AFM) phase, 
now canted in $x$ direction, weakens upon doping from half-filling, however at 
$n$$\sim$1.76 the Hubbard bilayer system shows a first-order phase transition to a 
metallic spin-flop (SF) phase. 
\begin{figure}[t]
\centering
\includegraphics*[width=6.5cm]{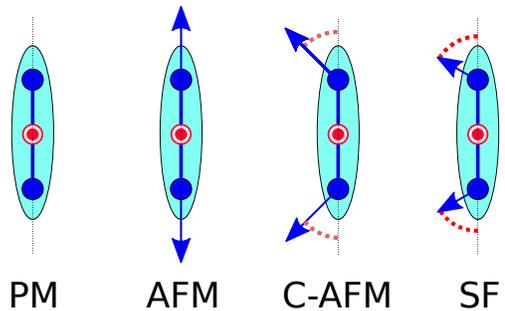}%
\caption{(Color online) Illustration of the stable local spin configurations
according to the local spin-spin angle $\gamma$ on the inter-layer cluster.
(a) PM without ordered local moments, (b) AFM with $\gamma$=$180^\circ$, 
(c) C-AFM with $\gamma$ between $180^\circ$ 
(pure AFM ordering) and $90^\circ$ (pure DM ordering), (d) SF with 
$\gamma$$<$$90^\circ$, i.e., weak ferromagnetism with strong canting.
\label{fig:hd2}}
\end{figure}
The latter one is characterized by the discontinuous jump to a local configuration
with an $\langle S_x\rangle$ expectation value {\sl larger} than 
$\langle S_z\rangle$. This corresponds to an angle $\gamma$ between the local spins 
in both layers being lower than $90^\circ$, whereas in the C-AFM phase 
$\gamma\in [90^\circ,180^\circ]$ holds (see Fig~\ref{fig:hd2}). The strong 
decrease of $\gamma$ at the transition point may be observed in the inset of 
Fig.~\ref{fig:hd1}. Hence the SF phase 
displays weak ferromagnetism due to strong canting. Note that neither the spin 
correlation functions nor the diagonal $Z$ values show a strong signature therein. 
The SF phase transforms into the usual PM phase at $n$$\sim$1.62. 
\begin{figure}[b]
\centering
\includegraphics*[width=8.5cm]{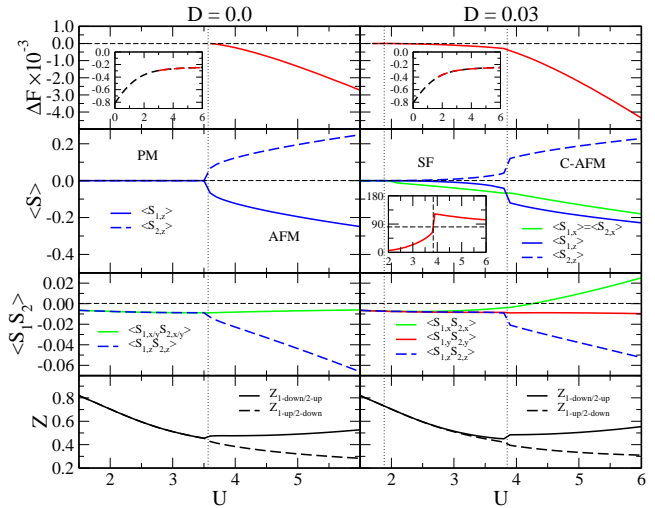}%
\caption{(Color online) Phase diagram with $U$ for the doped Hubbard bilayer at
filling $n$=1.7 (compare with Fig.~\ref{fig:hd1}). Insets show again the free-energy 
curves and the interaction-dependent spin-spin angle $\gamma$.\label{fig:hd7}}
\end{figure}

In addition to the doping scan, Fig.~\ref{fig:hd7} displays the various phases 
emerging with increasing Hubbard interaction $U$ for fixed hole doping 
$\delta$=0.3, i.e., $n$=1.7. Without the DM interaction, the standard picture of a
stable PM phase at small $U$ and a stable AFM phase at larger $U$ ($U$$>$3.58) 
remains vital. Note that the $U$ values for AFM stabilization are well above the 
Mott critical $U$ at half filling. Introducing $D$ stabilizes the metallic SF phase 
for 1.9$<$$U$$<$3.85, accompanied with the jump in the angle $\gamma$ towards 
lower values. Therewith the onset of AFM order takes place at slightly 
larger $U$ than for $D$=0. Hence the finite $D$ enables specific magnetic ordering 
in a Coulomb interacting regime that is originally not susceptible to such order. 
Only the $z$ component of the spin correlation function shows a discontinuous 
behavior at the SF/C-AFM phase boundary.

\section{ Two-impurity Anderson model\label{sec:tiam}}
The TIAM~\cite{jay82,fye87,jonkot89,fye89,schi96,nis06,ferr09} belongs to the set of
canonical models in the physics of strong electronic correlations, believed to be
relevant for the understanding of heavy-fermion systems~\cite{don77}. Via the
coupling of the impurities to a bath it contains the single-impurity Kondo physics 
and as competitor also the RKKY mechanism acting between the impurities. The latter 
originates from the effective exchange introduced through the impurity-coupling to 
the same bath. In some works~\cite{moe99,ferr09} this type of exchange interaction 
between sites is discussed in the context of two impurities coupled to different 
baths (similar to the bilayer architecture). But here we try to separate the 
exchange in an indirect ("RKKY") one, stemming from effective exchange via the bath,
and a direct term, resulting e.g. from an explicit hopping amplitude $t_{12}$ 
between the impurities (see Eq.~(\ref{eq:impham})).

\begin{figure}[b]
\centering
\includegraphics*[width=8.5cm]{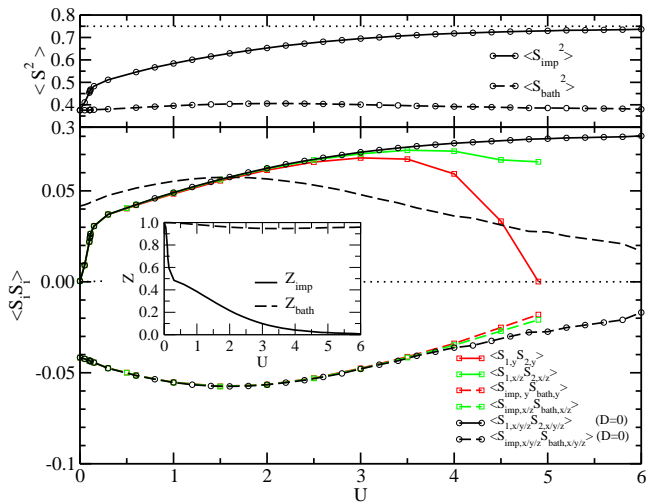}%
\caption{(Color online) Interaction-dependent spin correlation functions for the 
TIAM with $t_{12}$=0. Top: $\langle S^2\rangle$ for $D$=0, bottom: spin-spin
correlations for $D$=0 (circles) and $D$=0.05 (squares). The dashed dark line 
without circles is just the $\langle S_{\rm imp}S_{\rm bath}\rangle$ correlation 
function mirrored at the zero line. \label{fig:taim1}}
\end{figure}

Figure~\ref{fig:taim1} shows the RISB results for the spin correlation functions
of the fundamental model with $t_{12}$=$D$=0. Thus the two impurities 
are only coupled via the bath and exchange can only be mediated therewith.
The expectation value $\langle S^2\rangle$=$S(S+1)$ on the impurity quickly rises 
with $U$ due to the formation of the local moment. It approaches the value 3/4, 
corresponding to the full $S$=1/2 limit, at large interaction strength. With 
increasing $U$ a local Fermi liquid is established with a small quasiparticle 
weight $Z_{\rm imp}$ (see inset Fig.~\ref{fig:taim1}). The competition
between the Kondo screening and the RKKY interaction may be observed from 
inspection of the spin-spin correlations. From Fig.~\ref{fig:taim1} it is obvious
that $\langle S_{\rm imp}S_{\rm bath}\rangle$, i.e. the correlation between 
a single impurity and the bath, is always of AFM character with a maximum close to 
$U_{\rm K}$$\sim$1.6. On the other hand the inter-impurity correlation
$\langle S_1S_2\rangle$ is exclusively of FM kind and shows monotonic increase 
with $U$. The former is associated with the singlet-forming tendencies due to Kondo 
screening, whereas the latter signals triplet-forming tendencies because of the FM 
RKKY exchange within the local limit. Close to 
$U_{\rm K}$ the absolute value of the local RKKY correlation exceeds the 
singlet-forming amplitude between impurity and bath. The system at larger $U$ is 
then dominated by the RKKY interaction.~\cite{fye87,nis06,ferr09} Within a 
conventional Schrieffer-Wolff mapping~\cite{schr66} for the Kondo coupling via 
$J_{\rm K}$=$8V^2/U$, a similar crossover regime would follow also from simple 
estimates through the associated exchange interactions. For if we understand the 
RKKY interaction as second-order process, i.e. $J_{\rm RKKY}$$\sim$$J_{\rm K}^2$,
then here the two exchange integrals become equivalent for $U$=2, which is the 
order of magnitude from the numerics. With increasing impurity-bath coupling
$V$ the crossover shifts to larger $U$, since $J_{\rm K}^2$ stronger profits 
therefrom. However note that with our bath bandwidth $W$=6 the present TIAM is 
surely not in the Kondo-Hamiltonian limit ($U$$\gg$$W$) for the studied interaction 
range.~\cite{fye89}
\begin{figure}[t]
\centering
\includegraphics*[width=8.5cm]{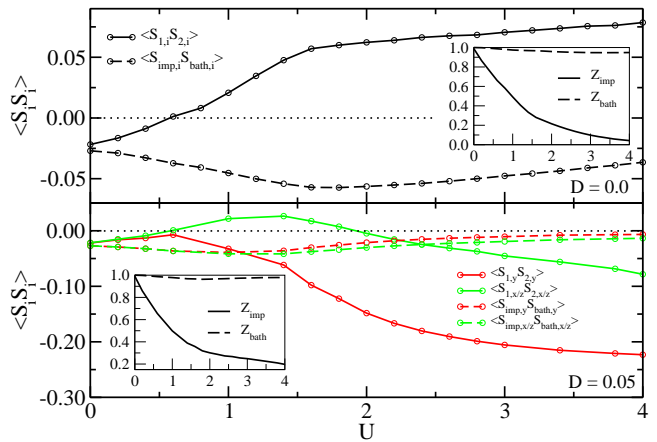}%
\caption{(Color online) Interaction-dependent spin-spin correlations for the TIAM 
with $t_{12}$=0.2, with and without DM interaction.
\label{fig:taim2}}
\end{figure}
Turning on a finite DM term of size $D$=0.05 has nearly no effect at small $U$. However
for larger Hubbard interaction rather strong modifications occur especially for the
inter-dimer function $\langle S_{1,y}S_{2,y}\rangle$. Remember that the $D$ vector 
also points in the $y$ direction. Thus an intricate spin-spin coupling scenario
arises at large $U$, with still FM alignment in the $(x,z)$ axes and near AFM
alignment in the $y$ axes. For $U$$>$5 our mean-field approach yields net local 
moments in presence of a finite $D$, i.e., a paramagnetic solution is no longer 
stabilizable. It would thus be very interesting to study the large-$U$ regime of this 
model beyond mean field (e.g. with the numerical renormalization group approach utilized
in Ref.~\onlinecite{nis06}).

In addition to the basic model with vanishing inter-impurity hopping, 
Fig.~\ref{fig:taim2} exhibits the resulting spin correlation functions for the TIAM 
with $t_{12}$=0.2. Now both $\langle S_{\rm imp}S_{\rm bath}\rangle$ and
$\langle S_1S_2\rangle$ display AFM correlations in the weakly interacting limit.
This is understood from the direct exchange integral $J_{\rm dir}$=$4t_{12}^2/U$
originating from the introduced dimer coupling. With increasing $U$ the correlation
functions develop rather similar as for $t_{12}$=0, yet the overall magnitude is
somewhat reduced a small interaction strength. Hence there the direct exchange weakens 
both, impurity Kondo-screening (due to the stronger inter-impurity link) as well as FM 
RKKY interaction (since the direct exchange favors AFM behavior). But the crossover 
point of domination for these processes does not seem to change much with the 
introduced $t_{12}$. Of course, a very large $t_{12}$ should rank the direct exchange 
above the other mechanisms, however here we do not investigate this model limit. 
Finally, when introducing the DM term to the model, effectively four different exchange 
mechanisms compete with each other: impurity Kondo, RKKY, direct and DM. The latter has 
indeed again significant effect on the spin correlation between the impurities. For 
already moderate values of $U$ the dominance of the FM RKKY is lost, turning the
system into AFM-like inter-impurity correlations for $U$$>$2. Thus also here the 
DM interaction severely influences the magnetic correlations for isolated impurities 
within an itinerant background. It appears to strengthen the singlet-forming tendencies 
(with stronger response in the ${\bf D}$ direction) in an otherwise triplet-favoring 
RKKY system at short-range distance.

\section{Summary}
A theoretical investigation of effects stemming from the Dzyaloshinskii-Moriya 
interaction in 
itinerant systems with strong electronic correlations was presented in this work.
In order to study the principle physics on the lattice as well as in the local 
limit, we elaborated on two prominent model systems, namely the Hubbard bilayer and
the one defined by the two-impurity Anderson Hamiltonian. In both cases substantial
influence of the DM interaction is found, especially at large coupling where the
strong renormalization enhances the impact. The half-filled Hubbard bilayer displays
large out-of-axis spin components close to the Mott transition that may severely 
influence the magnetic response in applied field. Intriguing phenomena in this
respect are e.g. observed in quasi-two-dimensional organic compounds.~\cite{kag08}
At finite hole doping and larger $U$, the bilayer system with DM interaction 
exhibits the emergence of a metallic spin-flop phase inbetween the AFM phase at 
half filling and the PM phase at stronger doping. This finding is of vital importance
for many doped Mott systems with anisotropies. For instance, it is well-known that
the hole-doped layered cuprates display puzzling phases inbetween the AFM and the 
superconducting dome and that the DM interaction is not completely negligible at
low energy.~\cite{thi88,cof91,jur06} Thus it would be very interesting to investigate
in some detail whether there is a closer connection between our model results and 
those observations.

The TIAM is very relevant not only in the context of heavy fermions, but e.g. also 
for isolated atoms on metallic surfaces. In either case, anisotropic spin terms 
such as the DM interaction exist in many realistic representants in nature.
It results from our studies that the DM term becomes an important player in the
hierachy of relevant exchange processes in these contexts. In the local limit
it works against the FM tendencies of the RKKY interaction and promotes the 
singlet formation between the impurities at large local Coloumb interactions.
Further research along these lines, e.g. by going beyond mean-field, including the 
complete $k$ dependence of the impurity-bath coupling or tailoring the modeling 
towards concrete materials systems is of vital interest to account for generic 
exchange processes in the strongly correlated itinerant regime.

\begin{acknowledgments}
We thank M.~Potthoff and D.~Grieger for helpful discussions.
Financial support from the Free and Hanseatic City of Hamburg in the context of the 
Landesexzellenzinitiative Hamburg as well as the DFG-SPP 1386 is gratefully 
acknowledged. Computations were performed at the local computing center of the 
University of Hamburg as well as the North-German Supercomputing Alliance (HLRN) 
under the grant hhp00026.
\end{acknowledgments}

\bibliography{bibextra}

\end{document}